\documentclass[reprint,aps,prl,amsmath,amssymb,showkeys]{revtex4-2}
\usepackage{graphicx}
\usepackage{bm}
\usepackage{mathptmx}
\usepackage{epstopdf}
\usepackage{booktabs}
\usepackage{multirow}
\usepackage{hhline}
\usepackage{subfigure}
\usepackage{amsmath}
\usepackage{mathrsfs}
\usepackage{amssymb}
\usepackage[colorlinks=true, allcolors=blue]{hyperref}
\newcommand{\hbf}{\mathbf{h}}
\newcommand{\sbf}{\mathbf{s}}
\newcommand{\ebf}{\mathbf{e}}
\newcommand{\zbf}{\mathbf{z}}
\newcommand{\ubf}{\mathbf{u}}
\newcommand{\wbf}{\mathbf{w}}
\newcommand{\cbf}{\mathbf{c}}
\newcommand{\vbf}{\mathbf{v}}
\newcommand{\Wbf}{\mathbf{W}}
\newcommand{\Vm}{\mathcal{V}_m}
\newcommand{\Vl}{\mathcal{V}_\ell}
\newcommand{\Gcal}{\mathcal{G}}
\newcommand{\Ecal}{\mathcal{E}}

\newcommand{\sigp}{\sigma_+}   
\newcommand{\sigi}{\sigma}     
\newcommand{\cat}{\,\|\,}      

\begin{document}

\title{A Magnetism-Aware Crystal Graph Neural Network with Explicit Magnetic-Sublattice and Exchange-Geometry Representations}
\author{Sourav Mal$^{1,2}$, Satadeep Bhattacharjee$^3$}
\affiliation{$^1$Harish-Chandra Research Institute, Chhatnag Road, Jhunsi, Prayagraj, 211019, India}
\affiliation{$^2$Homi Bhabha National Institute, Training School Complex, Anushakti Nagar, Mumbai, 400094, India}
\affiliation{$^3$Indo-Korea Science and Technology Center (IKST), Bengaluru 560065, India}
\email{s.bhattacharjee@ikst.res.in}

\begin{abstract}
Magnetic order in crystals is governed by moment-carrying sublattices and ligand-mediated exchange pathways, yet standard crystal graph neural networks do not explicitly distinguish magnetic sublattices and encode bonds primarily through pair distances. We propose mCGCNN, a magnetism-aware crystal graph neural network that augments the full structural graph with a dedicated magnetic subgraph. The magnetic stream performs angle-aware message passing between magnetic centers using metal--ligand--metal exchange-path descriptors motivated by Goodenough--Kanamori--Anderson physics, while layer-wise cross-coupling transfers structural and ligand-environment information from the full crystal graph. Separate pooling of the magnetic sublattice preserves magnetism-specific information alongside the global structural representation. We evaluate mCGCNN for total magnetic moment prediction on 49,462 crystalline materials with ferromagnetic spin configurations from the Materials Project and find that the proposed model improves agreement with DFT magnetic moments over established crystal-graph approaches without requiring a large increase in model complexity. Material-level analysis shows that the largest gains are concentrated predominantly in complex transition-metal and oxygen-containing magnetic compounds. Our results show that explicitly incorporating magnetic sublattices and exchange geometry provides a physically motivated route to improved crystal-graph representations for magnetic materials.
\end{abstract}

\maketitle

\textit{Introduction}---Accurate prediction of materials properties directly from crystal structure, without computationally expensive quantum-mechanical calculations, is a central goal of data-driven materials science. Graph neural networks (GNNs) have emerged as a powerful framework for this task, owing to their ability to naturally encode the topology and geometry of crystalline environments through iterative message passing over atom-centered graphs.

One of the earliest graph neural networks to demonstrate accurate materials-property prediction, the crystal graph convolutional neural network (CGCNN)~\cite{xie2018cgcnn} represents a crystal structure as a graph $\Gcal=(\mathcal{V},\mathcal{E})$, with atoms as nodes and neighboring atoms within a cutoff radius as edges. Atomic features are iteratively updated by
aggregating messages that encode the neighboring atomic features and interatomic distances, and the resulting atom-level representations are pooled into a crystal-level descriptor for property prediction.

Subsequent crystal-GNNs have extended CGCNN toward richer chemical and geometric representations. Extensions such as iCGCNN~\cite{iCGCNN} and MEGNet~\cite{MEGNet} enrich atomic, bonding, and global-state representations, while ALIGNN~\cite{choudhary2021alignn}, DimeNet~\cite{klicpera2020dimenet}, SchNet~\cite{schutt2018schnet}, and PaiNN~\cite{schutt2021painn} incorporate increasingly explicit geometric or directional information. Other approaches, such as CrysXPP~\cite{Das2022}, address limited labeled data through pretraining on large crystal databases. Nevertheless, conventional crystal GNNs generally lack explicit representations of magnetic sublattices and the ligand-mediated geometries that govern magnetic interactions.

Magnetic-property prediction presents several challenges for conventional crystal graph representations. The relevant information is concentrated on magnetic centers and their local exchange environments, which are not explicitly distinguished in the conventional CGCNN representation. All atomic sites are propagated through a common message-passing scheme, with their chemical identities encoded through the initial node features, and the resulting representations are pooled over the entire crystal. For a unit cell containing $N$ atoms but only $N_m\ll N$ magnetic sites, this pooling can dilute site-specific magnetic information through contributions from the
nonmagnetic majority. Moreover, CGCNN encodes interatomic interactions primarily through pairwise distances using Gaussian radial-basis functions
(RBFs)~\cite{xie2018cgcnn}, without explicitly representing bond angles. This is restrictive for ligand-mediated magnetic interactions, where the
geometry of the $\mathrm{M}$--$\mathrm{X}$--$\mathrm{M}$ exchange pathway can
substantially influence magnetic exchange; within the
Goodenough--Kanamori--Anderson framework~\cite{goodenough1958,kanamori1959,anderson1950}, for example, near-$180^\circ$ and near-$90^\circ$ pathways can favor different exchange tendencies.

Motivated by these limitations, we introduce mCGCNN, a dual-stream, angle-aware crystal graph neural network that explicitly distinguishes the magnetic sublattice from the full crystal environment and encodes the geometry of metal--ligand--metal exchange pathways. A dedicated magnetic subgraph, layer-wise coupling between structural and magnetic representations, and separate magnetic-sublattice pooling provide magnetism-specific inductive biases while retaining the efficiency of the crystal graph framework. Although the proposed framework is agnostic to magnetic ordering and can in principle be applied to different magnetic configurations, a systematic benchmark across magnetic ground states is
currently limited by the available data. We therefore evaluate the present model on 49,462 crystalline materials with ferromagnetic (FM) configurations
from the Materials Project~\cite{materials_project}, providing a controlled benchmark for total magnetic moment prediction and comparison with established crystal-graph neural networks.

\begin{figure}[htbp]
\centering
\includegraphics[width=0.48\textwidth]{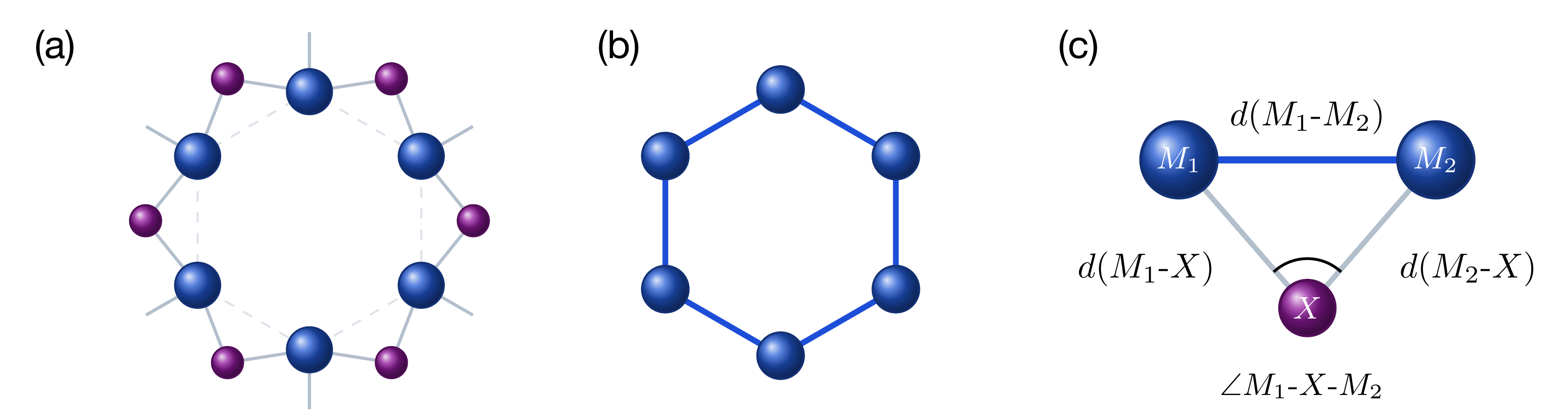}
\caption{Dual graph representation used in the mCGCNN architecture.
(a) The complete structural graph $\mathcal{G}$ incorporating all atoms in
the unit cell, with blue nodes denoting magnetic metal atoms and purple nodes
denoting non-magnetic/bridging atoms. (b) The corresponding magnetic
subgraph $\mathcal{G}_m$ consisting only of magnetic metal atoms. (c) The
magnetic edge features, comprising the magnetic-metal pair distance
$d(M_1-M_2)$, metal--bridge distances $d(M_1-X)$ and $d(M_2-X)$, the
exchange-path angle $\angle M_1-X-M_2$, and the features of the bridging
species $X$.}
\label{fig:magnetic_graph}
\end{figure}

\textit{Graph Representation}---Given a crystal structure with $N$ atoms per unit cell, two graphs are constructed: a structural graph $\Gcal$, identical to that used in the original CGCNN, and a magnetic subgraph $\Gcal_m$ constructed from candidate magnetic sites, as illustrated schematically in Fig.~\ref{fig:magnetic_graph}. The structural graph \(\mathcal{G}=(\mathcal{V},\mathcal{E})\) contains all \(N\) atoms as nodes, with edges $(i,j)\in\mathcal{E}$ connecting each atom to its neighboring atoms within a cutoff radius \(r_c^{\mathrm{str}}\), with the neighbor list padded to a fixed maximum of \(M\) neighbors. The edge feature vector for $(i,j)$ is constructed from the interatomic distance using a Gaussian RBF expansion, as discussed in the supplementary information (SI). The magnetic node set is defined as $
\Vm =
\left\{
i\in\mathcal{V}: Z_i\in\mathcal{Z}_m
\right\},
$ where $\mathcal{Z}_m$ denotes the set of atomic numbers corresponding to transition-metal, lanthanoid, and actinoid elements. Sites occupied by these elements are therefore designated as magnetic nodes, providing a chemistry-based proxy for the magnetic sublattice without assuming that every
selected site carries a finite magnetic moment. This criterion is used solely to construct the magnetic graph and does not rely on the target magnetic moment. The magnetic subgraph $\Gcal_m=(\Vm,\Ecal_m)$ connects pairs of magnetic nodes within a cutoff radius $r_c^{\mathrm{mag}}$, with the neighbor list padded to a fixed maximum of $M_m$ neighbors. The remaining atoms, $\Vl=\mathcal{V}\setminus\Vm$, are treated as non-magnetic sites and may serve as bridging ligands along magnetic exchange pathways. Their local chemical environments are therefore incorporated into the magnetic representation through the geometry and chemical character of the corresponding metal--ligand--metal pathways.

For each directed magnetic edge $(i,j)\in\Ecal_m$, the shortest bridging path through a non-magnetic site is identified as
$k^*=\arg\min_{k\in\Vl}\left(r_{ik}+r_{kj}\right)$, with the corresponding exchange-path angle given by
$
\cos\theta_{ik^*j}
=
\hat{\mathbf r}_{k^*i}\cdot\hat{\mathbf r}_{k^*j}$. For structures without a bridging non-magnetic site, such as elemental metals and transition-metal alloys, the corresponding M--M--M angle is used instead. The magnetic edge feature is then defined as
\begin{equation}
\ebf_{ij}^{\mathrm{mag}}
=
\bm{\phi}(r_{ij})
\cat
\bm{\phi}(r_{ik^*})
\cat
\bm{\phi}(r_{k^*j})
\cat
\bm{\psi}(\theta)
\cat
\mathbf e_{X_{k^*}},
\label{eq:mag_bond}
\end{equation}
where $\bm{\phi}$ denotes the radial basis for distance encoding, $\bm{\psi}(\theta)=[\cos\theta,\cos2\theta,\ldots,\cos K\theta]$ is a Fourier angular basis, and $\mathbf e_{X_{k^*}}$ is a one-hot encoding of the bridging-ligand species. Here, $\theta$ denotes the corresponding M--X--M or M--M--M angle. The resulting feature dimension is $d_b^{\mathrm{mag}}=3d_b+K+n_\ell$; for all-magnetic structures, the ligand encoding is replaced by an embedding of the central magnetic neighbor.

Each atom $i$ in the structural graph is initialized with the elemental feature vector $\vbf_i\in\mathbb{R}^{d_0}$ used in CGCNN, comprising period, group, electronegativity, covalent radius, valence-electron count, first ionization energy, electron affinity, block, and atomic volume. For magnetic nodes, this representation is augmented with seven magnetism-relevant elemental descriptors: atomic number, $d$- and
$f$-electron counts, unpaired $d$- and $f$-electron counts, and the spin and orbital angular momenta $(S,L)$ derived from Hund's rules.

\begin{figure*}[htbp]
\centering
\includegraphics[width=1\textwidth]{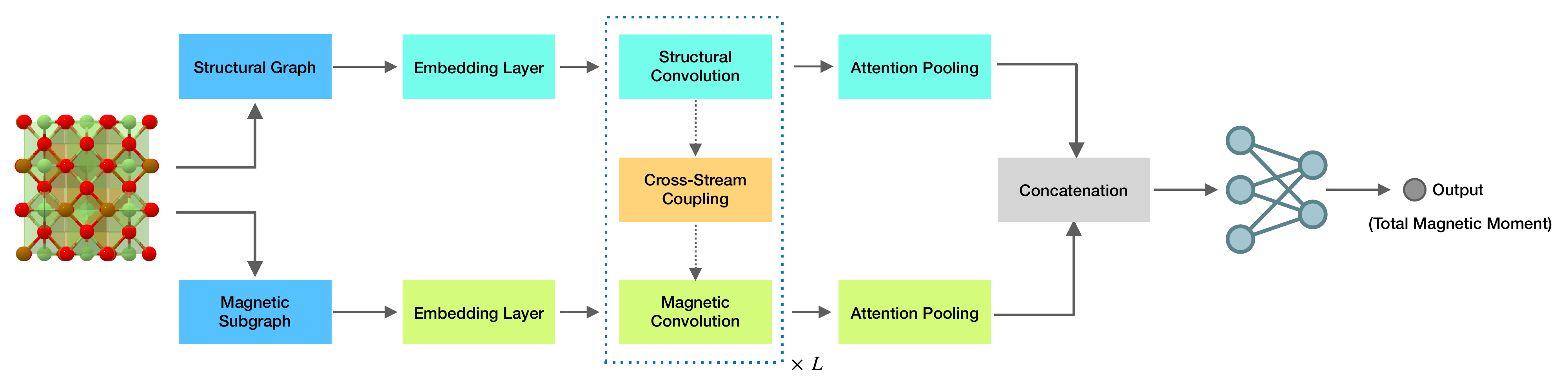}
\caption{Schematic architecture of mCGCNN. The structural stream processes the complete crystal graph, while the magnetic stream operates on the magnetic subgraph using angle-aware gated convolutions. At each convolutional layer, structural representations at magnetic sites are transferred to the
magnetic stream through cross-stream coupling. The two streams are separately pooled by attention and concatenated to form the crystal representation $\cbf_{\mathrm{crys}}$, which is mapped to the predicted total magnetic moment through a fusion and residual MLP head.}
\label{fig:mcgcnn_architecture}
\end{figure*}

\textit{Architecture}---The proposed mCGCNN architecture, illustrated in
Fig.~\ref{fig:mcgcnn_architecture}, comprises two parallel streams that encode the full structural graph and the corresponding magnetic subgraph, respectively. The two streams are initialized by separate embedding layers that project
their input node features into independent latent spaces:
\begin{align}
  \hbf_i^{(0)}
  &=
  \Wbf_{\mathrm{emb}}^{\mathrm{str}}\,\vbf_i
  + \mathbf{b}_{\mathrm{emb}}^{\mathrm{str}},
  \qquad
  \hbf_i^{(0)}\in\mathbb{R}^{d_a},
  \quad i\in\mathcal{V},\\
  \sbf_i^{(0)}
  &=
  \Wbf_{\mathrm{emb}}^{\mathrm{mag}}\,\vbf_i
  + \mathbf{b}_{\mathrm{emb}}^{\mathrm{mag}},
  \qquad
  \sbf_i^{(0)}\in\mathbb{R}^{d_s},
  \quad i\in\Vm,
\end{align}
where $d_a$ and $d_s$ denote the hidden dimensions of the structural and
magnetic streams, respectively, and are chosen independently.

The structural and magnetic streams are propagated in lockstep through $L$ convolutional layers. At each layer $\ell$, the structural stream updates all atomic representations using the gated convolution adopted from CGCNN. For $i\in\mathcal{V}$ and $j\in\mathcal{N}(i)$, the edge message is formed as
\begin{equation}
\mathbf{m}_{ij}
=
\hbf_i^{(\ell)}
\cat
\hbf_j^{(\ell)}
\cat
\bm{\phi}(r_{ij}),
\end{equation}
and transformed into filter and core components,
\begin{equation}
\zbf_{ij}
=
\mathrm{BN}_1\!\left(
\Wbf_{\mathrm{str}}^\top\mathbf{m}_{ij}
+\mathbf{b}_{\mathrm{str}}
\right),
\end{equation}
where $\zbf_{ij}^{(f)}$ and $\zbf_{ij}^{(c)}$ denote the corresponding filter and core components. The node representation is then updated by the gated aggregation
\begin{equation}
\hbf_i^{(\ell+1)}
=
\sigma_{+}\!\left[
\hbf_i^{(\ell)}
+
\mathrm{BN}_2\!\left(
\sum_{j\in\mathcal{N}(i)}
\sigma\!\left(\zbf_{ij}^{(f)}\right)
\odot
\sigma_{+}\!\left(\zbf_{ij}^{(c)}\right)
\right)
\right].
\label{eq:str_update}
\end{equation} Here, $\sigma$ and $\sigma_{+}$ denote the sigmoid and softplus activation functions, respectively, and $\mathrm{BN}$ denotes batch normalization.

The updated structural representation is subsequently coupled to the
magnetic stream at the magnetic sites $\mathcal{V}_m$ through a learned,
layer-specific projection,
$
\sbf_i^{(\ell)}
\leftarrow
\sbf_i^{(\ell)}
+
\Wbf_{\mathrm{cross}}^{(\ell)}
\hbf_i^{(\ell+1)},~i\in\mathcal{V}_m$,
where $\Wbf_{\mathrm{cross}}^{(\ell)}\in\mathbb{R}^{d_s\times d_a}$.
The coupling is unidirectional, allowing the magnetic representation to
receive structural and ligand-environment information from the full
crystal graph without modifying the structural stream. This transfer is
important because the local chemical environment surrounding a magnetic
center provides information about the ligand-mediated exchange pathway.
For example, Fe--O--Fe and Fe--F--Fe environments can involve similar
magnetic centers while possessing distinct ligand-mediated orbital
environments. The cross-coupling therefore connects the magnetic
representation to the structural context required to distinguish such
exchange pathways.

The magnetic stream then performs gated message passing on the magnetic
subgraph using the angle-aware exchange descriptors
$\ebf_{ij}^{\mathrm{mag}}$ introduced above. For magnetic neighbors
$j\in\mathcal{N}_m(i)$, the message is $\wbf_{ij}
=\sbf_i^{(\ell)}\cat\sbf_j^{(\ell)}\cat\ebf_{ij}^{\mathrm{mag}}$, and the magnetic representation is updated as
\begin{align}
\ubf_{ij}
&=
\mathrm{BN}_1^m\!\left(
\Wbf_{\mathrm{mag}}^\top\wbf_{ij}
+\mathbf{b}_{\mathrm{mag}}
\right),\\
\sbf_i^{(\ell+1)}
&=
\sigp\!\left[
\sbf_i^{(\ell)}
+
\mathrm{BN}_2^m\!\left(
\sum_{j\in\mathcal{N}_m(i)}
\sigi(\ubf_{ij}^{(f)})
\odot
\sigp(\ubf_{ij}^{(c)})
\right)
\right],
\label{eq:mag_update}
\end{align}
where $\ubf_{ij}^{(f)}$ and $\ubf_{ij}^{(c)}$ are the filter and core components of $\ubf_{ij}$. 
Unlike the structural stream, the magnetic convolution operates only on magnetic centers and incorporates the geometry of their exchange pathways. After each structural update, the representations of the magnetic sites are extracted and transformed by a layer-specific bias-free linear map, then added residually to the magnetic stream before its convolutional update. Equations~\eqref{eq:str_update}--\eqref{eq:mag_update} therefore define one coupled magnetism-aware convolutional layer, repeated for $L$ layers. This layer-wise coupling combines the broader crystal environment with site-selective magnetic connectivity and exchange geometry.

After $L$ rounds of coupled message passing, the structural and magnetic streams are pooled separately using additive attention to obtain crystal-level representations. For $q\in\{\mathrm{str},\mathrm{mag}\}$,
the attention weights are defined as
\begin{equation}
\alpha_i^{(q)}
=
\frac{
\exp\!\left[
\mathbf{w}_q^\top
\tanh\!\left(
\Wbf_q\mathbf{h}_i^{(L)}
\right)
\right]
}{
\sum_{j\in\mathcal{V}_q}
\exp\!\left[
\mathbf{w}_q^\top
\tanh\!\left(
\Wbf_q\mathbf{h}_j^{(L)}
\right)
\right]
},
\end{equation}
where $\mathcal{V}_{\mathrm{str}}=\mathcal{V}$ and
$\mathcal{V}_{\mathrm{mag}}=\mathcal{V}_m$. The corresponding pooled representations are
\begin{equation}
\cbf_q
=
\sum_{i\in\mathcal{V}_q}
\alpha_i^{(q)}\mathbf{h}_i^{(L)},
\qquad
q\in\{\mathrm{str},\mathrm{mag}\},
\end{equation}
and the crystal descriptor is obtained by concatenation,
\begin{equation}
\cbf_{\mathrm{crys}}
=
\cbf_{\mathrm{str}}\cat\cbf_{\mathrm{mag}}
\in\mathbb{R}^{d_a+d_s}.
\label{eq:crystal_descriptor}
\end{equation}
For structures without identified magnetic centers, the magnetic
descriptor is set to the zero vector. The resulting crystal descriptor is then passed to the prediction head.

The crystal representation $\cbf_{\mathrm{crys}}$ is mapped to the
prediction space by a fusion layer followed by $n_h$ residual blocks:
\begin{equation}
\mathbf{c}^{(0)}
=
\mathrm{Drop}\!\left[
\sigma_+\!\left(
\Wbf_{\mathrm{fuse}}\,
\mathrm{LN}(\cbf_{\mathrm{crys}})
+\mathbf{b}_{\mathrm{fuse}}
\right)
\right],
\label{eq:fusion}
\end{equation}
where $\mathrm{Drop}$ denotes a dropout layer with rate $p=0.1$, and
$\Wbf_{\mathrm{fuse}}\in\mathbb{R}^{h_{\mathrm{fea}}\times(d_a+d_s)}$.
For $k=1,\ldots,n_h$, the residual blocks are
\begin{align}
\mathbf{z}^{(k)}
&=
\Wbf_k\mathbf{c}^{(k-1)}+\mathbf{b}_k,\\
\mathbf{c}^{(k)}
&=
\mathbf{c}^{(k-1)}
+
\mathrm{Drop}\!\left[
\sigma_+\!\left(
\mathrm{LN}(\mathbf{z}^{(k)})
\right)
\right],
\label{eq:residual}
\end{align}
with $\Wbf_k\in\mathbb{R}^{h_{\mathrm{fea}}\times h_{\mathrm{fea}}}$.
The output is
\begin{equation}
\hat{y}
=
\Wbf_{\mathrm{out}}\mathbf{c}^{(n_h)}
+\mathbf{b}_{\mathrm{out}},
\label{eq:output}
\end{equation}
where $\hat{y}$ is the predicted total magnetic moment per unit cell. Detailed model hyperparameters (Table S1), training procedures, and computational complexity relative to CGCNN are provided in the SI.

\begin{table}[htbp]
    \centering
    \caption{Performance comparison of CGCNN, MEGNet, ALIGNN, and mCGCNN for
    total magnetic moment prediction on the test set. All models are trained and
    evaluated on the same data splits from the Materials Project FM
    configurations. ``Params'' denotes the number of trainable parameters,
    reported in millions.}
    
    \renewcommand{\arraystretch}{1.2} 
    \setlength{\tabcolsep}{12pt}      
    
    \begin{tabular}{lcccc}
        \toprule
        Model & Params & MAE & RMSE & R$^2$ \\
        \midrule
        CGCNN  & 0.08 & 5.477 & 9.763 & 0.639 \\
        MEGNet & 0.19 & 5.236 & 9.904 & 0.628 \\
        ALIGNN & 4.03 & 3.653 & 8.654 & 0.716 \\
        mCGCNN & 0.37 & 4.344 & 7.910 & 0.763 \\
        \bottomrule
    \end{tabular}
    \label{tab:model_performance}
\end{table}

\begin{figure}[htbp]
\centering
\includegraphics[width=0.45\textwidth]{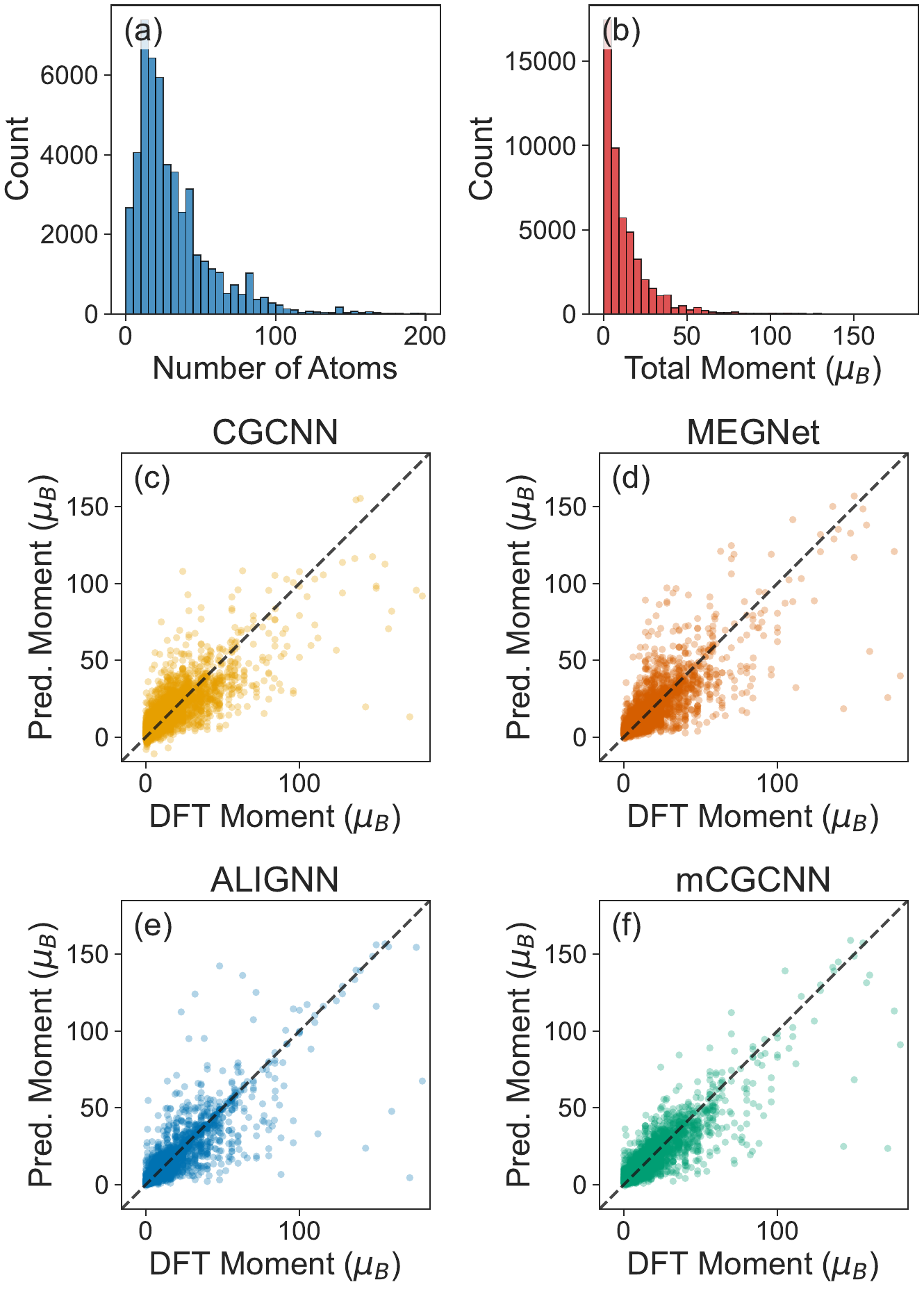}
\caption{Dataset distributions and predictive performance for total magnetic
moment. Top row: distributions of (a) the number of atoms per unit cell and
(b) the DFT-calculated total magnetic moment ($\mu_B$) across the benchmark
dataset. Bottom rows: parity plots of predicted versus DFT-calculated magnetic
moments for (c) CGCNN, (d) MEGNet, (e) ALIGNN, and (f) the proposed mCGCNN
architecture. The dashed line denotes perfect agreement ($y=x$).}
\label{fig:parity_plots}
\end{figure}

\textit{Results}---We trained mCGCNN on a curated dataset of 49,462 crystalline materials with ferromagnetic (FM) spin configurations from the Materials Project to predict the total magnetic moment per unit cell. To assess the performance of the proposed representation, CGCNN, MEGNet, and ALIGNN were trained and evaluated on the same dataset using identical training, validation, and test splits. The dataset contains 49,462 inorganic crystalline materials with ferromagnetic (FM) spin configurations, with 1--392 atoms per unit cell (mean: 31). The atom-count distribution is shown in Fig.~\ref{fig:parity_plots}(a), restricted to structures with at most 200 atoms for clarity. The total magnetic moments have a mean of 13.29~$\mu_B$ per cell and are strongly concentrated at lower values, with an approximately exponential decrease in density toward larger moments [Fig.~\ref{fig:parity_plots}(b)].

The dataset spans diverse chemical environments, with transition-metal oxides comprising 36.67\%, followed by oxypnictides (17.52\%), oxyhalides (9.57\%), transition-metal halides (4.99\%), transition-metal intermetallics (4.51\%), transition-metal chalcogenides (4.41\%),
oxychalcogenides (3.93\%), and other mixed-anion compounds (3.55\%); the remainder comprises pnictides, nitrides, carbides, borides, silicides, hydrides, and rare-earth-containing compounds.

We evaluate the four models on the held-out test set using the mean absolute error (MAE), root-mean-square error (RMSE), and coefficient of determination ($R^2$). As shown in Table~\ref{tab:model_performance}, mCGCNN outperforms
CGCNN and MEGNet across all three metrics, reducing the MAE from 5.477 and 5.236 to 4.344~$\mu_B$, the RMSE from 9.763 and 9.904 to 7.910~$\mu_B$, and increasing $R^2$ from 0.639 and 0.628 to 0.763, respectively. ALIGNN achieves the lowest MAE (3.653~$\mu_B$), but has a higher RMSE (8.654~$\mu_B$) and lower $R^2$ (0.716) than mCGCNN. The parity plots in Fig.~\ref{fig:parity_plots}(c--f) are consistent with these
metrics: mCGCNN exhibits comparatively tighter agreement with the ideal $y=x$ line and fewer pronounced deviations at large magnetic moments, consistent with its lower RMSE and higher $R^2$.

An important aspect of this comparison is the relationship between predictive performance and model complexity. The four models contain approximately $0.08$, $0.19$, $0.37$, and $4.0$ million trainable parameters for CGCNN, MEGNet, mCGCNN, and ALIGNN, respectively. Despite having only a moderate increase in model size over CGCNN and MEGNet, mCGCNN achieves a lower RMSE and higher $R^2$ than ALIGNN, while remaining competitive in MAE, despite having roughly an order of magnitude fewer trainable parameters. This comparison suggests that increasing architectural complexity alone is not necessarily the most effective route to improved magnetic-property prediction. Instead, explicitly incorporating physically motivated inductive biases, here through magnetic-sublattice structure and exchange-path geometry, can provide substantial predictive gains without requiring a substantial increase in model complexity.

To identify the chemical environments in which mCGCNN provides the largest
material-level advantage, we examined compositions for which it is the most
accurate of the four models (Tables~S2 and S3). The corresponding winning
cases are predominantly transition-metal-containing oxygen-rich compounds,
including complex oxides and mixed-anion oxides. Among materials for which
ALIGNN is the strongest baseline (Table~S2), few of the representative examples include
$\mathrm{MnCrCoO_4}$, $\mathrm{VO_2}$, $\mathrm{Fe_2O_3}$, $\mathrm{Fe_3O_4}$,
$\mathrm{Sr_{15}Fe_{10}(Bi_5O_{23})_2}$, and $\mathrm{FeO_2F}$. For
$\mathrm{Sr_{15}Fe_{10}(Bi_5O_{23})_2}$, mCGCNN reduces the absolute error
relative to ALIGNN by $28.03~\mu_B$ (from $69.56$ to $41.53~\mu_B$ for a
DFT value of $96.00~\mu_B$), while reductions of $23.48$ and
$23.11~\mu_B$ are obtained for $\mathrm{MnCrCoO_4}$ and $\mathrm{VO_2}$,
respectively. mCGCNN also remains the most accurate model when CGCNN or
MEGNet provides the strongest baseline (Table~S3), with representative
examples including $\mathrm{Mn_3Cr_3O_8}$, $\mathrm{LiFeO_2}$, and
$\mathrm{SrLa_3Mn_3CrO_{12}}$. For the high-moment
$\mathrm{Mn_2AsHO_5}$, mCGCNN reduces the error from $78.18$ to
$23.69~\mu_B$, a reduction of $54.50~\mu_B$. Although such improvements
also occur in halides, pnictides, chalcogenides, borides, and intermetallics,
the winning cases are predominantly concentrated in the broader chemical
space of complex transition-metal and oxygen-containing magnetic materials.

\textit{Discussion}---We have introduced mCGCNN, a magnetism-aware crystal graph framework that explicitly separates the magnetic sublattice from the
full crystal structure and encodes the local geometry of exchange pathways. By coupling structural and magnetic representations layer by layer, the
model retains the broader chemical environment while focusing message passing and pooling on magnetic centers and their surrounding exchange geometry. This provides a physically motivated inductive bias for learning magnetic properties directly from crystal structure while retaining the compact graph-based formulation of conventional crystal GNNs.

More broadly, our results show that incorporating magnetism-specific structure into crystal representations can improve the description of
magnetic materials without requiring a fundamentally different graph learning framework. The present formulation provides a foundation for extending crystal-graph models toward richer magnetic descriptors, including site-resolved moments, magnetic ordering temperatures, and non-collinear magnetic configurations. Such extensions, together with larger datasets containing systematically characterized magnetic ground states and multiple magnetic solutions, may further broaden the scope of
magnetism-aware materials prediction.

More generally, this work illustrates how physically and chemically motivated inductive biases can be incorporated into existing crystal graph
neural networks without substantially altering their underlying framework. The same design principle may be applicable beyond magnetism, including to
other classes of complex materials such as energy-storage materials and quantum materials, where identifying and explicitly encoding property-relevant local structural or chemical environments may provide
similarly useful inductive biases.

\textit{Acknowledgements}---S.~Mal acknowledges the high-performance computing facility~\cite{hri_hpc} at the Harish-Chandra Research Institute (HRI), Prayagraj, India, for providing the computational resources used in this work. This work was supported by KIST (Korea Institute of Science and Technology) Institutional Program: Global Knowledge Platform Research Project (26Z0100).

\textit{Code availability}---The code is openly available in Ref.~\cite{code_repo}. 

\bibliography{references}

\begin{thebibliography}{14}%
\makeatletter
\providecommand \@ifxundefined [1]{%
 \@ifx{#1\undefined}
}%
\providecommand \@ifnum [1]{%
 \ifnum #1\expandafter \@firstoftwo
 \else \expandafter \@secondoftwo
 \fi
}%
\providecommand \@ifx [1]{%
 \ifx #1\expandafter \@firstoftwo
 \else \expandafter \@secondoftwo
 \fi
}%
\providecommand \natexlab [1]{#1}%
\providecommand \enquote  [1]{``#1''}%
\providecommand \bibnamefont  [1]{#1}%
\providecommand \bibfnamefont [1]{#1}%
\providecommand \citenamefont [1]{#1}%
\providecommand \href@noop [0]{\@secondoftwo}%
\providecommand \href [0]{\begingroup \@sanitize@url \@href}%
\providecommand \@href[1]{\@@startlink{#1}\@@href}%
\providecommand \@@href[1]{\endgroup#1\@@endlink}%
\providecommand \@sanitize@url [0]{\catcode `\\12\catcode `\$12\catcode
  `\&12\catcode `\#12\catcode `\^12\catcode `\_12\catcode `\%12\relax}%
\providecommand \@@startlink[1]{}%
\providecommand \@@endlink[0]{}%
\providecommand \url  [0]{\begingroup\@sanitize@url \@url }%
\providecommand \@url [1]{\endgroup\@href {#1}{\urlprefix }}%
\providecommand \urlprefix  [0]{URL }%
\providecommand \Eprint [0]{\href }%
\providecommand \doibase [0]{https://doi.org/}%
\providecommand \selectlanguage [0]{\@gobble}%
\providecommand \bibinfo  [0]{\@secondoftwo}%
\providecommand \bibfield  [0]{\@secondoftwo}%
\providecommand \translation [1]{[#1]}%
\providecommand \BibitemOpen [0]{}%
\providecommand \bibitemStop [0]{}%
\providecommand \bibitemNoStop [0]{.\EOS\space}%
\providecommand \EOS [0]{\spacefactor3000\relax}%
\providecommand \BibitemShut  [1]{\csname bibitem#1\endcsname}%
\let\auto@bib@innerbib\@empty
\bibitem [{\citenamefont {Xie}\ and\ \citenamefont
  {Grossman}(2018)}]{xie2018cgcnn}%
  \BibitemOpen
  \bibfield  {author} {\bibinfo {author} {\bibfnamefont {T.}~\bibnamefont
  {Xie}}\ and\ \bibinfo {author} {\bibfnamefont {J.~C.}\ \bibnamefont
  {Grossman}},\ }\bibfield  {title} {\bibinfo {title} {Crystal graph
  convolutional neural networks for an accurate and interpretable prediction of
  material properties},\ }\href
  {https://doi.org/10.1103/PhysRevLett.120.145301} {\bibfield  {journal}
  {\bibinfo  {journal} {Physical Review Letters}\ }\textbf {\bibinfo {volume}
  {120}},\ \bibinfo {pages} {145301} (\bibinfo {year} {2018})}\BibitemShut
  {NoStop}%
\bibitem [{\citenamefont {Park}\ and\ \citenamefont
  {Wolverton}(2020)}]{iCGCNN}%
  \BibitemOpen
  \bibfield  {author} {\bibinfo {author} {\bibfnamefont {C.~W.}\ \bibnamefont
  {Park}}\ and\ \bibinfo {author} {\bibfnamefont {C.}~\bibnamefont
  {Wolverton}},\ }\bibfield  {title} {\bibinfo {title} {Developing an improved
  crystal graph convolutional neural network framework for accelerated
  materials discovery},\ }\href
  {https://doi.org/10.1103/PhysRevMaterials.4.063801} {\bibfield  {journal}
  {\bibinfo  {journal} {Phys. Rev. Mater.}\ }\textbf {\bibinfo {volume} {4}},\
  \bibinfo {pages} {063801} (\bibinfo {year} {2020})}\BibitemShut {NoStop}%
\bibitem [{\citenamefont {Chen}\ \emph {et~al.}(2019)\citenamefont {Chen},
  \citenamefont {Ye}, \citenamefont {Zuo}, \citenamefont {Zheng},\ and\
  \citenamefont {Ong}}]{MEGNet}%
  \BibitemOpen
  \bibfield  {author} {\bibinfo {author} {\bibfnamefont {C.}~\bibnamefont
  {Chen}}, \bibinfo {author} {\bibfnamefont {W.}~\bibnamefont {Ye}}, \bibinfo
  {author} {\bibfnamefont {Y.}~\bibnamefont {Zuo}}, \bibinfo {author}
  {\bibfnamefont {C.}~\bibnamefont {Zheng}},\ and\ \bibinfo {author}
  {\bibfnamefont {S.~P.}\ \bibnamefont {Ong}},\ }\bibfield  {title} {\bibinfo
  {title} {Graph networks as a universal machine learning framework for
  molecules and crystals},\ }\href
  {https://doi.org/10.1021/acs.chemmater.9b01294} {\bibfield  {journal}
  {\bibinfo  {journal} {Chemistry of Materials}\ }\textbf {\bibinfo {volume}
  {31}},\ \bibinfo {pages} {3564} (\bibinfo {year} {2019})},\ \Eprint
  {https://arxiv.org/abs/https://doi.org/10.1021/acs.chemmater.9b01294}
  {https://doi.org/10.1021/acs.chemmater.9b01294} \BibitemShut {NoStop}%
\bibitem [{\citenamefont {Choudhary}\ and\ \citenamefont
  {DeCost}(2021)}]{choudhary2021alignn}%
  \BibitemOpen
  \bibfield  {author} {\bibinfo {author} {\bibfnamefont {K.}~\bibnamefont
  {Choudhary}}\ and\ \bibinfo {author} {\bibfnamefont {B.}~\bibnamefont
  {DeCost}},\ }\bibfield  {title} {\bibinfo {title} {Atomistic line graph
  neural network for improved materials property predictions},\ }\href@noop {}
  {\bibfield  {journal} {\bibinfo  {journal} {npj Computational Materials}\
  }\textbf {\bibinfo {volume} {7}},\ \bibinfo {pages} {185} (\bibinfo {year}
  {2021})}\BibitemShut {NoStop}%
\bibitem [{\citenamefont {Gasteiger}\ \emph {et~al.}(2020)\citenamefont
  {Gasteiger}, \citenamefont {Gro{\ss}},\ and\ \citenamefont
  {G{\"u}nnemann}}]{klicpera2020dimenet}%
  \BibitemOpen
  \bibfield  {author} {\bibinfo {author} {\bibfnamefont {J.}~\bibnamefont
  {Gasteiger}}, \bibinfo {author} {\bibfnamefont {J.}~\bibnamefont
  {Gro{\ss}}},\ and\ \bibinfo {author} {\bibfnamefont {S.}~\bibnamefont
  {G{\"u}nnemann}},\ }\bibfield  {title} {\bibinfo {title} {Directional message
  passing for molecular graphs},\ }\href@noop {} {\bibfield  {journal}
  {\bibinfo  {journal} {arXiv preprint arXiv:2003.03123}\ } (\bibinfo {year}
  {2020})}\BibitemShut {NoStop}%
\bibitem [{\citenamefont {Schutt}\ \emph {et~al.}(2018)\citenamefont {Schutt},
  \citenamefont {Sauceda}, \citenamefont {Kindermans}, \citenamefont
  {Tkatchenko},\ and\ \citenamefont {Müller}}]{schutt2018schnet}%
  \BibitemOpen
  \bibfield  {author} {\bibinfo {author} {\bibfnamefont {K.~T.}\ \bibnamefont
  {Schutt}}, \bibinfo {author} {\bibfnamefont {H.~E.}\ \bibnamefont {Sauceda}},
  \bibinfo {author} {\bibfnamefont {P.-J.}\ \bibnamefont {Kindermans}},
  \bibinfo {author} {\bibfnamefont {A.}~\bibnamefont {Tkatchenko}},\ and\
  \bibinfo {author} {\bibfnamefont {K.-R.}\ \bibnamefont {Müller}},\
  }\bibfield  {title} {\bibinfo {title} {Schnet – a deep learning
  architecture for molecules and materials},\ }\href@noop {} {\bibfield
  {journal} {\bibinfo  {journal} {The Journal of Chemical Physics}\ }\textbf
  {\bibinfo {volume} {148}},\ \bibinfo {pages} {241722} (\bibinfo {year}
  {2018})}\BibitemShut {NoStop}%
\bibitem [{\citenamefont {Sch{\"u}tt}\ \emph {et~al.}(2021)\citenamefont
  {Sch{\"u}tt}, \citenamefont {Unke},\ and\ \citenamefont
  {Gastegger}}]{schutt2021painn}%
  \BibitemOpen
  \bibfield  {author} {\bibinfo {author} {\bibfnamefont {K.}~\bibnamefont
  {Sch{\"u}tt}}, \bibinfo {author} {\bibfnamefont {O.}~\bibnamefont {Unke}},\
  and\ \bibinfo {author} {\bibfnamefont {M.}~\bibnamefont {Gastegger}},\
  }\bibfield  {title} {\bibinfo {title} {Equivariant message passing for the
  prediction of tensorial properties and molecular spectra},\ }in\ \href@noop
  {} {\emph {\bibinfo {booktitle} {International conference on machine
  learning}}}\ (\bibinfo {organization} {PMLR},\ \bibinfo {year} {2021})\ pp.\
  \bibinfo {pages} {9377--9388}\BibitemShut {NoStop}%
\bibitem [{\citenamefont {Das}\ \emph {et~al.}(2022)\citenamefont {Das},
  \citenamefont {Samanta}, \citenamefont {Goyal}, \citenamefont {Lee},
  \citenamefont {Bhattacharjee},\ and\ \citenamefont {Ganguly}}]{Das2022}%
  \BibitemOpen
  \bibfield  {author} {\bibinfo {author} {\bibfnamefont {K.}~\bibnamefont
  {Das}}, \bibinfo {author} {\bibfnamefont {B.}~\bibnamefont {Samanta}},
  \bibinfo {author} {\bibfnamefont {P.}~\bibnamefont {Goyal}}, \bibinfo
  {author} {\bibfnamefont {S.-C.}\ \bibnamefont {Lee}}, \bibinfo {author}
  {\bibfnamefont {S.}~\bibnamefont {Bhattacharjee}},\ and\ \bibinfo {author}
  {\bibfnamefont {N.}~\bibnamefont {Ganguly}},\ }\bibfield  {title} {\bibinfo
  {title} {Crysxpp: An explainable property predictor for crystalline
  materials},\ }\href {https://doi.org/10.1038/s41524-022-00716-8} {\bibfield
  {journal} {\bibinfo  {journal} {npj Computational Materials}\ }\textbf
  {\bibinfo {volume} {8}},\ \bibinfo {pages} {43} (\bibinfo {year}
  {2022})}\BibitemShut {NoStop}%
\bibitem [{\citenamefont {Goodenough}(1958)}]{goodenough1958}%
  \BibitemOpen
  \bibfield  {author} {\bibinfo {author} {\bibfnamefont {J.~B.}\ \bibnamefont
  {Goodenough}},\ }\bibfield  {title} {\bibinfo {title} {An interpretation of
  the magnetic properties of the perovskite-type mixed crystals la1- xsrxcoo3-
  $\lambda$},\ }\href@noop {} {\bibfield  {journal} {\bibinfo  {journal}
  {Journal of Physics and chemistry of Solids}\ }\textbf {\bibinfo {volume}
  {6}},\ \bibinfo {pages} {287} (\bibinfo {year} {1958})}\BibitemShut {NoStop}%
\bibitem [{\citenamefont {Kanamori}(1959)}]{kanamori1959}%
  \BibitemOpen
  \bibfield  {author} {\bibinfo {author} {\bibfnamefont {J.}~\bibnamefont
  {Kanamori}},\ }\bibfield  {title} {\bibinfo {title} {Superexchange
  interaction and symmetry properties of electron orbitals},\ }\href
  {https://doi.org/10.1016/0022-3697(59)90061-7} {\bibfield  {journal}
  {\bibinfo  {journal} {Journal of Physics and Chemistry of Solids}\ }\textbf
  {\bibinfo {volume} {10}},\ \bibinfo {pages} {87} (\bibinfo {year}
  {1959})}\BibitemShut {NoStop}%
\bibitem [{\citenamefont {Anderson}(1950)}]{anderson1950}%
  \BibitemOpen
  \bibfield  {author} {\bibinfo {author} {\bibfnamefont {P.~W.}\ \bibnamefont
  {Anderson}},\ }\bibfield  {title} {\bibinfo {title} {Antiferromagnetism.
  theory of superexchange interaction},\ }\href
  {https://doi.org/10.1103/PhysRev.79.350} {\bibfield  {journal} {\bibinfo
  {journal} {Physical Review}\ }\textbf {\bibinfo {volume} {79}},\ \bibinfo
  {pages} {350} (\bibinfo {year} {1950})}\BibitemShut {NoStop}%
\bibitem [{\citenamefont {Horton}\ \emph {et~al.}(2025)\citenamefont {Horton},
  \citenamefont {Huck}, \citenamefont {Yang}, \citenamefont {Munro},
  \citenamefont {Dwaraknath}, \citenamefont {Ganose}, \citenamefont
  {Kingsbury}, \citenamefont {Wen}, \citenamefont {Shen}, \citenamefont
  {Mathis}, \citenamefont {Kaplan}, \citenamefont {Berket}, \citenamefont
  {Riebesell}, \citenamefont {George}, \citenamefont {Rosen}, \citenamefont
  {Spotte-Smith}, \citenamefont {McDermott}, \citenamefont {Cohen},
  \citenamefont {Dunn}, \citenamefont {Kuner}, \citenamefont {Rignanese},
  \citenamefont {Petretto}, \citenamefont {Waroquiers}, \citenamefont
  {Griffin}, \citenamefont {Neaton}, \citenamefont {Chrzan}, \citenamefont
  {Asta}, \citenamefont {Hautier}, \citenamefont {Cholia}, \citenamefont
  {Ceder}, \citenamefont {Ong}, \citenamefont {Jain},\ and\ \citenamefont
  {Persson}}]{materials_project}%
  \BibitemOpen
  \bibfield  {author} {\bibinfo {author} {\bibfnamefont {M.~K.}\ \bibnamefont
  {Horton}}, \bibinfo {author} {\bibfnamefont {P.}~\bibnamefont {Huck}},
  \bibinfo {author} {\bibfnamefont {R.~X.}\ \bibnamefont {Yang}}, \bibinfo
  {author} {\bibfnamefont {J.~M.}\ \bibnamefont {Munro}}, \bibinfo {author}
  {\bibfnamefont {S.}~\bibnamefont {Dwaraknath}}, \bibinfo {author}
  {\bibfnamefont {A.~M.}\ \bibnamefont {Ganose}}, \bibinfo {author}
  {\bibfnamefont {R.~S.}\ \bibnamefont {Kingsbury}}, \bibinfo {author}
  {\bibfnamefont {M.}~\bibnamefont {Wen}}, \bibinfo {author} {\bibfnamefont
  {J.~X.}\ \bibnamefont {Shen}}, \bibinfo {author} {\bibfnamefont {T.~S.}\
  \bibnamefont {Mathis}}, \bibinfo {author} {\bibfnamefont {A.~D.}\
  \bibnamefont {Kaplan}}, \bibinfo {author} {\bibfnamefont {K.}~\bibnamefont
  {Berket}}, \bibinfo {author} {\bibfnamefont {J.}~\bibnamefont {Riebesell}},
  \bibinfo {author} {\bibfnamefont {J.}~\bibnamefont {George}}, \bibinfo
  {author} {\bibfnamefont {A.~S.}\ \bibnamefont {Rosen}}, \bibinfo {author}
  {\bibfnamefont {E.~W.~C.}\ \bibnamefont {Spotte-Smith}}, \bibinfo {author}
  {\bibfnamefont {M.~J.}\ \bibnamefont {McDermott}}, \bibinfo {author}
  {\bibfnamefont {O.~A.}\ \bibnamefont {Cohen}}, \bibinfo {author}
  {\bibfnamefont {A.}~\bibnamefont {Dunn}}, \bibinfo {author} {\bibfnamefont
  {M.~C.}\ \bibnamefont {Kuner}}, \bibinfo {author} {\bibfnamefont {G.-M.}\
  \bibnamefont {Rignanese}}, \bibinfo {author} {\bibfnamefont {G.}~\bibnamefont
  {Petretto}}, \bibinfo {author} {\bibfnamefont {D.}~\bibnamefont
  {Waroquiers}}, \bibinfo {author} {\bibfnamefont {S.~M.}\ \bibnamefont
  {Griffin}}, \bibinfo {author} {\bibfnamefont {J.~B.}\ \bibnamefont {Neaton}},
  \bibinfo {author} {\bibfnamefont {D.~C.}\ \bibnamefont {Chrzan}}, \bibinfo
  {author} {\bibfnamefont {M.}~\bibnamefont {Asta}}, \bibinfo {author}
  {\bibfnamefont {G.}~\bibnamefont {Hautier}}, \bibinfo {author} {\bibfnamefont
  {S.}~\bibnamefont {Cholia}}, \bibinfo {author} {\bibfnamefont
  {G.}~\bibnamefont {Ceder}}, \bibinfo {author} {\bibfnamefont {S.~P.}\
  \bibnamefont {Ong}}, \bibinfo {author} {\bibfnamefont {A.}~\bibnamefont
  {Jain}},\ and\ \bibinfo {author} {\bibfnamefont {K.~A.}\ \bibnamefont
  {Persson}},\ }\bibfield  {title} {\bibinfo {title} {Accelerated data-driven
  materials science with the materials project},\ }\href
  {https://doi.org/10.1038/s41563-025-02272-0} {\bibfield  {journal} {\bibinfo
  {journal} {Nature Materials}\ }\textbf {\bibinfo {volume} {24}},\ \bibinfo
  {pages} {1522} (\bibinfo {year} {2025})}\BibitemShut {NoStop}%
\bibitem [{hri()}]{hri_hpc}%
  \BibitemOpen
  \href@noop {} {}\bibinfo {howpublished}
  {\url{https://www.hri.res.in/cluster/}}\BibitemShut {NoStop}%
\bibitem [{cod()}]{code_repo}%
  \BibitemOpen
  \href@noop {} {}\bibinfo {howpublished}
  {\url{https://github.com/SouravMal/mCGCNN}}\BibitemShut {NoStop}%
\end{thebibliography}%
\clearpage
\section{Supplementary Information\\ A Magnetism-Aware Crystal Graph Neural Network with Explicit Magnetic-Sublattice and Exchange-Geometry Representations}
\author{Sourav Mal$^{1,2}$, Satadeep Bhattacharjee$^3$}
\affiliation{$^1$Harish-Chandra Research Institute, Chhatnag Road, Jhunsi, Prayagraj, 211019, India}
\affiliation{$^2$Homi Bhabha National Institute, Training School Complex, Anushakti Nagar, Mumbai, 400094, India}
\affiliation{$^3$Indo-Korea Science and Technology Center (IKST), Bengaluru 560065, India}
\maketitle

\section{Radial Basis Function Expansion}

Interatomic distances are encoded using a Gaussian radial basis expansion,
following the original CGCNN formulation. For an interatomic distance
$r_{ij}$, the radial feature vector is defined as
\begin{equation}
  \bm{\phi}(r_{ij}) =
  \left[
    \exp\!\left(
      -\frac{(r_{ij}-\mu_k)^2}{2\sigma^2}
    \right)
  \right]_{k=1}^{d_b}
  \in \mathbb{R}^{d_b},
\end{equation}
where $d_b=41$ Gaussian basis functions have centers $\{\mu_k\}$ uniformly
spaced over the interval $[0,r_c^{\mathrm{str}}]$, with width
$\sigma=0.2$~\AA. The same radial basis expansion is used for the
interatomic distances appearing in the structural graph.

\section{Computational Complexity}
The additional computational cost of mCGCNN over CGCNN is bounded by
the magnetic subgraph operations.  Per convolutional layer, the magnetic
stream requires $O(N_m \cdot M_m \cdot d_s^2)$ floating-point operations,
compared to $O(N \cdot M \cdot d_a^2)$ for the structural stream.  Since
$N_m \leq N$ and typically $M_m \leq M$, the magnetic stream never
exceeds the structural stream in cost.  For systems where $N_m \ll N$
(dilute magnetic systems), the total cost is dominated by the structural
stream and the overhead of mCGCNN over CGCNN is negligible.  The
cross-coupling projection adds $O(N_m \cdot d_a \cdot d_s)$ operations
per layer, which is sub-dominant under the same conditions.

\section{Training Details}

The dataset was split into training, validation, and test sets in an 80:10:10 ratio. The target magnetic moments are standardized using the mean and standard deviation computed exclusively from the training set. The normalized targets
are used during optimization and transformed back to the original units for
evaluation and prediction. The regression objective is based on the mean
squared error (MSE),
\begin{equation}
\mathcal{L}_{\mathrm{MSE}}
=
\frac{1}{N}
\sum_{i=1}^{N}
\left(
\hat{y}_i-y_i^{\mathrm{norm}}
\right)^2,
\end{equation}
where $y_i^{\mathrm{norm}}$ and $\hat{y}_i$ denote the normalized DFT and
predicted total magnetic moments, respectively.

For smaller training sets, $L_2$ regularization can be introduced
separately for the structural and magnetic parameter groups. The resulting
objective is
\begin{equation}
\mathcal{L}_{\mathrm{total}}
=
\mathcal{L}_{\mathrm{MSE}}
+
\lambda_{\mathrm{str}}
\sum_{\theta\in\Theta_{\mathrm{str}}}
\|\theta\|_2^2
+
\lambda_{\mathrm{mag}}
\sum_{\theta\in\Theta_{\mathrm{mag}}}
\|\theta\|_2^2,
\end{equation}
where $\Theta_{\mathrm{str}}$ and $\Theta_{\mathrm{mag}}$ denote the
trainable parameters of the structural and magnetic streams, respectively,
and $\lambda_{\mathrm{str}}$ and $\lambda_{\mathrm{mag}}$ are their
corresponding weight-decay coefficients. In particular, the latter can be
chosen larger when stronger regularization of the magnetic stream is desired.
Because the present dataset contains approximately $5\times10^4$ materials,
the final mCGCNN model was trained without weight decay, i.e.,
$\lambda_{\mathrm{str}}=\lambda_{\mathrm{mag}}=0$.

The model is optimized using the Adam optimizer with an initial learning rate
of $10^{-4}$. The learning rate is reduced by a factor of $0.1$ at epochs 80
and 120 using a multi-step learning-rate scheduler. Training is performed for
150 epochs with a batch size of 128. Dropout with probability $p=0.1$ is
applied in the prediction head, including the fusion layer and residual
multilayer perceptron blocks.

The model corresponding to the lowest validation MSE is retained as the best
checkpoint and subsequently evaluated on the held-out test set. All
experiments use a fixed random seed of 42, with deterministic CuDNN
operations enabled to ensure reproducibility. The model was trained on a single NVIDIA H100 GPU using 8 CPU cores and 32 GB of allocated host memory.

\begin{table}[h]
\centering
\caption{Hyperparameters used for training mCGCNN. Parameters introduced
specifically for the magnetism-aware stream are marked with $\star$.}
\label{tab:hyper}
\begin{tabular}{lll}
\toprule
Parameter & Symbol & Default \\
\midrule
Structural stream node feature dim        & $d_0$                  & 92 \\
Structural stream hidden dim        & $d_a$                  & 64 \\
Magnetic stream hidden dim$^\star$  & $d_s$                  & 64 \\
Number of convolution layers       & $L$                    & 3 \\
Structural Gaussian RBF centres     & $d_b^{\mathrm{str}}$   & 41 \\
Magnetic Gaussian RBF centres$^\star$ & $d_b^{\mathrm{mag}}$ & 26 \\
Fourier angle terms$^\star$         & $K$                    & 8 \\
Post-pool hidden dim                & $h_{\rm fea}$          & 256 \\
Residual blocks after pooling       & $n_h$                  & 2 \\
Structural cutoff                   & $r_c^{\rm str}$        & 8.0~\AA \\
Magnetic cutoff$^\star$             & $r_c^{\rm mag}$        & 5.0~\AA \\
Maximum structural neighbors        & $M$                    & 12 \\
Maximum magnetic neighbors$^\star$  & $M_m$                  & 8 \\
Dropout probability                 & $p$                    & 0.1 \\
Weight decay                        & $\lambda$              & 0 \\
Batch size                          & $B$                    & 128 \\
Learning rate                       & $\eta$                 & $10^{-4}$ \\
Optimizer                           & ---                    & Adam \\
\bottomrule
\end{tabular}
\end{table}

\begin{table*}[htbp]
    \centering
    \caption{Comparison of total magnetic moments predicted by ALIGNN and
mCGCNN for the top 32 materials exhibiting the largest predictive advantage of
mCGCNN over ALIGNN. The target values correspond to DFT-calculated total
magnetic moments per unit cell. $N_{\mathrm{sites}}$ denotes the number of
atomic sites in the unit cell. The advantage is defined as the reduction in
absolute prediction error obtained with mCGCNN relative to ALIGNN,
$\mathrm{AE}_{\mathrm{ALIGNN}}-\mathrm{AE}_{\mathrm{mCGCNN}}$, where
$\mathrm{AE}=|y_{\mathrm{pred}}-y_{\mathrm{DFT}}|$. Magnetic moments are
reported in $\mu_B$.}
    \label{tab:mcgcnn_vs_alignn}
    \begin{tabular}{llccccc}
        \toprule
        Material ID & Formula & $N_{\text{sites}}$ & Target & ALIGNN & mCGCNN & Advantage \\
        \midrule
        mp-aaabjcea & Sr$_{15}$Fe$_{10}$(Bi$_{5}$O$_{23}$)$_{2}$ & 162 & 96.00 & 41.53 & 69.56 & 28.03 \\
        mp-aaacrnlx & MnCrCoO$_{4}$ & 14 & 16.00 & 49.19 & 25.71 & 23.48 \\
        mp-aaacsvul & VO$_{2}$ & 102 & 36.00 & 12.15 & 35.26 & 23.11 \\
        mp-aaabrcrb & LiCoPO$_{4}$ & 168 & 72.06 & 37.64 & 86.87 & 19.61 \\
        mp-aaacsvux & VO$_{2}$ & 102 & 34.00 & 12.37 & 28.54 & 16.17 \\
        mp-aaabglmx & MnCoSb & 24 & 35.87 & 11.85 & 25.11 & 13.25 \\
        mp-aaabreoc & Li$_{4}$Mn$_{5}$(NiO$_{6}$)$_{2}$ & 46 & 46.00 & 22.96 & 36.19 & 13.23 \\
        mp-aaacsvtc & VO$_{2}$ & 102 & 34.00 & 10.67 & 23.05 & 12.39 \\
        mp-aaabsntf & Li$_{2}$Cr(FeO$_{3}$)$_{2}$ & 33 & 36.00 & 22.13 & 38.68 & 11.19 \\
        mp-aaacrfmw & Y$_{4}$ZrMn$_{5}$O$_{15}$ & 50 & 42.00 & 27.45 & 45.61 & 10.94 \\
        mp-aaabjmtq & EuAg & 8 & 32.00 & 19.09 & 35.01 & 9.90 \\
        mp-aaabscfh & NaV$_{5}O_{8}$ & 56 & 40.00 & 27.49 & 43.04 & 9.47 \\
        mp-aaabrepr & FeO$_{2}$F & 8 & 10.00 & 20.02 & 9.33 & 9.35 \\
        mp-aaacrruz & Gd$_{2}$CoB$_{3}$ & 6 & 14.52 & 4.36 & 12.28 & 7.92 \\
        mp-aaacqbmx & Tm$_{2}$CrS$_{4}$ & 56 & 31.75 & 15.94 & 23.25 & 7.31 \\
        mp-aaacrnie & Mn$_{7}$Sb$_{4}$ & 11 & 20.41 & 12.37 & 19.64 & 7.27 \\
        mp-aaabrhlw & Li$_{5}$V$_{3}$Cu(PO$_{4}$)$_{4}$ & 58 & 18.00 & 9.38 & 16.42 & 7.04 \\
        mp-aaabrhrz & Na$_{2}$V$_{13}$O$_{22}$ & 74 & 46.00 & 29.02 & 35.98 & 6.96 \\
        mp-aaacqtdd & V$_{2}$Cu$_{3}$O$_{11}$ & 16 & 1.41 & 11.98 & 5.02 & 6.96 \\
        mp-aaacqaxp & MnSb$_{2}$H$_{8}$(SN)$_{4}$ & 38 & 10.00 & 19.26 & 12.97 & 6.29 \\
        mp-aaacoytf & LiFe$_{2}$F$_{5}$ & 32 & 32.00 & 39.58 & 30.36 & 5.94 \\
        mp-aaacwiac & Li$_{2}$V$_{3}$WO$_{8}$ & 14 & 7.00 & 12.79 & 7.14 & 5.64 \\
        mp-aaadales & Fe$_{2}$O$_{3}$ & 20 & 40.00 & 47.03 & 38.60 & 5.62 \\
        mp-aaabosxf & Fe$_{3}O_{4}$ & 28 & 56.00 & 65.94 & 60.37 & 5.58 \\
        mp-aaacrmht & NaEuTiNbO$_{6}$F & 22 & 12.00 & 19.55 & 14.08 & 5.47 \\
        mp-aaabrflf & LiFe$_{4}$(PO$_{4}$)$_{3}$ & 80 & 64.00 & 49.70 & 73.03 & 5.27 \\
        mp-aaabnsms & Eu$_{6}$Mg$_{7}$H$_{26}$ & 39 & 42.00 & 27.39 & 32.52 & 5.13 \\
        mp-aaabfxwm & V$_{3}$SiP$_{5}$O$_{19}$ & 112 & 24.03 & 14.99 & 20.05 & 5.06 \\
        mp-aaabiduu & FeHC$_{2}$O$_{3}$ & 28 & 16.00 & 21.21 & 15.76 & 4.97 \\
        mp-aaabdbno & MnO$_{2}$ & 6 & 6.00 & 14.43 & 9.47 & 4.96 \\
        mp-aaacsvvw & Fe$_{2}$O$_{3}$ & 80 & 150.00 & 156.11 & 148.76 & 4.87 \\
        mp-aaabrypi & LiAlFeO$_{4}$ & 28 & 16.02 & 11.01 & 16.20 & 4.83 \\
        \bottomrule
    \end{tabular}
    \label{tab:alignn_advantage}
\end{table*}

\begin{table*}[htbp]
    \centering
    \caption{Comparison of total magnetic moments predicted by CGCNN, MEGNet,
ALIGNN, and mCGCNN for the top 30 materials exhibiting the largest predictive
advantage of mCGCNN over the baseline models. $N_{\mathrm{sites}}$ denotes
the number of atomic sites in the unit cell, and the target values correspond
to DFT-calculated total magnetic moments per unit cell. The ``Best Baseline''
column identifies the baseline model with the smallest absolute prediction
error for each material. The advantage is defined as the reduction in
absolute prediction error obtained with mCGCNN relative to the best-performing
baseline, $\Delta\mathrm{AE} =
\mathrm{AE}_{\mathrm{baseline}}-\mathrm{AE}_{\mathrm{mCGCNN}}$, where
$\mathrm{AE}=|y_{\mathrm{pred}}-y_{\mathrm{DFT}}|$. Magnetic moments are
reported in $\mu_B$.}
    \label{tab:mcgcnn_vs_all_baselines}
    \resizebox{\textwidth}{!}{
    \begin{tabular}{llcccccccc}
        \toprule
        Material ID & Formula & $N_{\text{sites}}$ & Target & Best Baseline & CGCNN & MEGNet & ALIGNN & mCGCNN & Advantage \\
        \midrule
        mp-aaabrocn & Mn$_{2}$AsHO$_{5}$ & 144 & 160.05 & CGCNN & 81.87 & 55.70 & 47.68 & 136.36 & 54.50 \\
        mp-aaacqwzx & Mn$_{13}$Fe(SiO$_{11}$)$_{2}$ & 38 & 63.00 & CGCNN & 108.20 & 120.89 & 136.16 & 66.20 & 42.00 \\
        mp-aaacrnke & Mn$_{3}$Cr$_{3}$O$_{8}$ & 14 & 23.00 & CGCNN & 54.23 & 86.21 & 112.36 & 22.41 & 30.63 \\
        mp-aaabwkdz & LiFeO$_{2}$ & 64 & 80.00 & MEGNet & 24.58 & 29.00 & 18.72 & 54.88 & 25.88 \\
        mp-aaabqxik & GdAsO$_{4}$ & 12 & 14.00 & CGCNN & 37.01 & 58.56 & 49.83 & 10.92 & 19.94 \\
        mp-aaabrdor & Li$_{4}$Fe$_{5}$(NiO$_{6}$)$_{2}$ & 46 & 56.00 & CGCNN & 28.28 & 27.72 & 26.45 & 47.96 & 19.69 \\
        mp-aaabntjh & Ca$_{4}$Mn$_{4}$O$_{11}$ & 38 & 28.00 & CGCNN & 65.00 & 97.00 & 95.00 & 45.81 & 19.19 \\
        mp-aaabsiwz & LiFeOF$_{2}$ & 40 & 40.00 & CGCNN & 23.27 & 21.09 & 20.32 & 39.67 & 16.40 \\
        mp-aaabrrub & Sr$_{16}$Mn$_{8}$O$_{29}$ & 53 & 30.00 & CGCNN & 13.93 & 12.60 & 11.89 & 29.04 & 15.10 \\
        mp-aaacqqfe & VCoO$_{3}$ & 10 & 8.00 & MEGNet & 31.52 & 26.25 & 44.97 & 11.35 & 14.90 \\
        mp-aaacqnnu & Gd(BO$_{2}$)$_{3}$ & 160 & 112.00 & CGCNN & 64.39 & 32.16 & 33.04 & 78.64 & 14.26 \\
        mp-aaacrvis & CaNdMn$_{2}$O$_{6}$ & 20 & 14.00 & CGCNN & 32.87 & 48.69 & 45.22 & 18.99 & 13.88 \\
        mp-aaabcsro & Fe$_{3}$BO$_{5}$ & 18 & 26.00 & MEGNet & 59.93 & 49.42 & 50.12 & 36.46 & 12.97 \\
        mp-aaacreye & Zn$_{5}$Cr$_{20}$(GaSe$_{10}$)$_{4}$ & 69 & 59.88 & MEGNet & 19.68 & 34.11 & 27.20 & 45.73 & 11.62 \\
        mp-aaabrwbs & LiCr$_{2}$O$_{4}$ & 28 & 20.00 & MEGNet & 36.97 & 32.09 & 34.92 & 20.59 & 11.50 \\
        mp-aaacryhc & Ba$_{3}$Eu$_{3}$(Fe$_{3}$O$_{8}$)$_{2}$ & 28 & 49.00 & MEGNet & 34.33 & 36.97 & 21.97 & 49.68 & 11.35 \\
        mp-aaabfrnz & MnBiO$_{3}$ & 40 & 32.00 & CGCNN & 43.15 & 17.75 & 18.29 & 31.74 & 10.88 \\
        mp-aaacriek & SrLa$_{3}$Mn$_{3}$CrO$_{12}$ & 20 & 14.00 & CGCNN & 26.76 & 27.79 & 34.64 & 16.02 & 10.75 \\
        mp-aaabrcob & Ti$_{2}$Fe$_{4}$O$_{9}$ & 30 & 36.00 & CGCNN & 51.87 & 58.23 & 59.94 & 30.65 & 10.53 \\
        mp-aaabrghe & Li$_{2}$Mn$_{2}$F$_{7}$ & 44 & 36.00 & CGCNN & 22.59 & 22.06 & 21.21 & 38.97 & 10.44 \\
        mp-aaacszok & SrCa$_{3}$Mn$_{4}$O$_{10}$ & 36 & 32.00 & MEGNet & 67.46 & 49.36 & 124.02 & 39.33 & 10.04 \\
        mp-aaaabmky & MnPO$_{4}$ & 12 & 8.00 & CGCNN & 30.63 & 31.85 & 35.89 & 20.75 & 9.88 \\
        mp-aaabwjlh & CoPO$_{4}$ & 144 & 95.99 & MEGNet & 28.46 & 44.17 & 27.23 & 53.84 & 9.67 \\
        mp-aaabrmym & Li$_{3}$Co$_{2}$(SiO$_{4}$)$_{2}$ & 60 & 28.00 & MEGNet & 14.51 & 15.50 & 9.05 & 24.94 & 9.43 \\
        mp-aaabsgco & Li$_{2}$FeF$_{4}$ & 56 & 32.00 & CGCNN & 18.68 & 11.95 & 15.73 & 28.01 & 9.33 \\
        mp-aaacrkou & Sr$_{10}$Mn$_{8}$B$_{2}$O$_{25}$ & 90 & 64.00 & MEGNet & 29.15 & 50.77 & 19.26 & 67.90 & 9.33 \\
        mp-aaabqymt & Li$_{2}$FeF$_{4}$ & 14 & 8.00 & CGCNN & 21.11 & 25.78 & 25.25 & 11.95 & 9.16 \\
        mp-aaacshsg & LiCr$_{2}$O$_{4}$ & 7 & 5.00 & MEGNet & 21.63 & 16.65 & 20.41 & 7.76 & 8.89 \\
        mp-aaaewfxa & MgFe$_{2}$(PO$_{4}$)$_{2}$ & 13 & 8.00 & CGCNN & 20.92 & 23.62 & 21.74 & 12.08 & 8.85 \\
        mp-aaabcscj & Cr$_{4}$(PO$_{4}$)$_{3}$ & 76 & 60.00 & CGCNN & 38.84 & 28.44 & 31.28 & 47.57 & 8.73 \\
        \bottomrule
    \end{tabular}
    }
    \label{tab:all_model_advantage}
\end{table*}

\end{document}